\documentclass[prl,aps,twocolumn]{revtex4}
\usepackage{epsfig}
\usepackage{bm}
\newcommand{\B}[1]{{\bm{#1}}}

\usepackage[latin1]{inputenc}
\newcommand{\beq}{\begin{equation}}
\newcommand{\eeq}{\end{equation}}
\newcommand{\bea}{\begin{eqnarray}}
\newcommand{\eea}{\end{eqnarray}}
\begin{document}
\title{Roughening of Fracture Surfaces: the Role of Plastic Deformations \\ Version of \today}
\author{Eran Bouchbinder$^1$, Joachim Mathiessen$^{1,2}$ and Itamar Procaccia$^1$}
\affiliation{$^1$Dept. of Chemical Physics, The Weizmann Institute
of Science, Rehovot 76100, Israel\\
 $^2$The Niels Bohr Institute, 17 Blegdamsvej, Copenhagen, Denmark}
\begin{abstract}

Post mortem analysis of fracture surfaces of ductile and brittle
materials on the $\mu$m-mm and the nm scales respectively, reveal
self affine graphs with an anomalous scaling exponent
$\zeta\approx 0.8$. Attempts to use elasticity theory to explain
this result failed,  yielding exponent $\zeta\approx 0.5$ up to
logarithms. We show that when the cracks propagate via plastic
void formations in front of the tip, followed by void coalescence,
the voids positions are positively correlated to yield exponents
higher than 0.5.
\end{abstract}
\maketitle

Quantitative studies of fracture surfaces reveal self affine rough
graphs with two scaling regimes: at small length scales (smaller
than a typical cross-over length $\xi_c$) the roughness exponents
$\zeta\approx 0.5$, whereas at scales larger than $\xi_c$ the
roughness exponent is $\zeta\approx 0.8$. Such measurements were
reported first for ductile materials (like metals) where $\xi_c$
is of the order of 1 $\mu$m \cite{84MPP, 97B}, and more recently
for brittle materials like glass, but with a much smaller value of
$\xi_c$, of about 1 nm \cite{03C}. The exponent  $\zeta\approx
0.5$ is characteristic of uncorrelated random walks, but higher
exponents indicate correlated steps \cite{Feder}; naturally the
experimental discovery of such correlated, ``anomalous" exponents
attracted considerable interest with repeated attempts to derive them
theoretically. Up to now these attempts were based on elasticity
theory and have failed to underpin the mechanism for correlated
fracture steps. For realistic boundary conditions, i.e. mode I or
mode II fracture, these attempts invariably ended up with logarithmic 
roughening \cite{97REF} or with the random walk scaling
exponents $\zeta\approx 0.5$ \cite{02BLP}.

In this Letter we present a quantitative model for ductile fracture in an infinite 2-dimensional
material that
follows the qualitative picture presented recently in \cite{99BP}, see Fig.1. In this picture there exists
a ``process zone" in front of the crack tip in which plastic yield is accompanied by the evolution of
damage cavities. A crucial aspect of this picture is the existence of a typical scale, $\xi_c$, which
is roughly the distance between the crack tip and the first void, at the time of the nucleation of the
latter. The voids are nucleated under the influence of the stress field $\sigma_{ij}(\B r)$ adjacent to the
tip, but not {\em at} the tip, due to the existence of the plastic zone that cuts off the purely
linear-elastic (unphysical) crack-tip singularities. The crack grows
by coalescing the voids with the tip, creating a new stress field which induces the
nucleation of new voids. In the picture of \cite{99BP} the scale $\xi_c$ is
also identified with the typical size of the voids at coalescence. A consequence of this picture
is that the roughening exponent $\zeta\approx 0.5$ corresponds to the surface structure of
individual voids, whereas the exponent $\zeta\approx 0.8$ has to do with the correlation between
the positions of different voids that coalesce to constitute the evolving crack. To dress this picture
with quantitative content we need first to provide a theory for the scale $\xi_c$ and, second,
to demonstrate that the positions of consecutive voids are positively correlated. These are the main
goals of this Letter.

%%%%%%%%%% FIGURE 1 %%%%%%%%%%%%
\begin{figure}
\centering
\epsfig{width=.45\textwidth,file=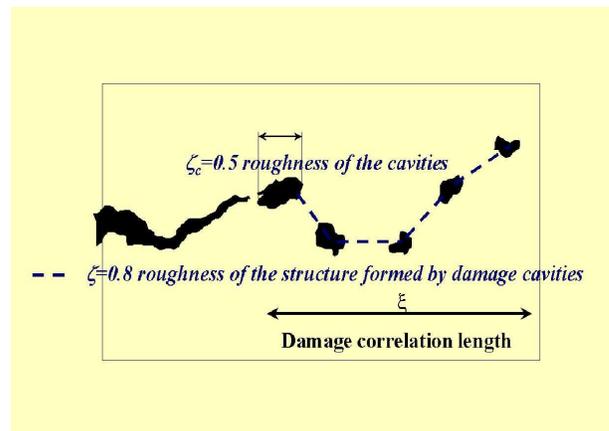}
\caption{The fracture scenario suggested in \cite{99BP}. This scenario had been documented in detail in corrosive glass
fracture, and also more recently in the fracture of paper \cite{03Cil}. Figure
courtesy of E. Bouchaud.}
\label{Yield}
\end{figure}
%%%%%%%%%%%%%%%%%%%%%%%%%%%%%%%

A simple model for $\xi_c$ can be developed by assuming the
process zone to be properly described by the Huber-von Mises
plasticity theory \cite{90Lub}. This theory focuses on the deviatoric stress
$s_{ij}\equiv \sigma_{ij} -\case{1}{3}{\rm Tr}
\B\sigma\delta_{ij}$ and on its invariants. The second invariant,
$J_2\equiv \case{1}{2} s_{ij}s_{ij}$, corresponds to the
distortional energy. The material yields as the distortional
energy exceeds a material-dependent threshold $\sigma_{_{\rm
Y}}^2$. In 2-dimensions this yield condition reads \cite{90Lub}
\begin{equation}
J_2 =\frac{\sigma^2_{1}-\sigma_{1}\sigma_{2}+\sigma^2_{2}}{3}=\sigma^2_{_{\rm Y}} \ .
\label{Mises}
\end{equation}
Here $\sigma_{1\!,2}$ are the principal stresses given by
\begin{equation}
\sigma_{1\!,2} = \frac{\sigma_{yy}+\sigma_{xx}}{2} \pm
\sqrt{\frac{(\sigma_{yy}-\sigma_{xx})^2}{4}+\sigma_{xy}^2} \ .
\label{PrincipleStress}
\end{equation}

In the purely linear-elastic solution the crack-tip region is
where high stresses are concentrated (in fact diverging near a
sharp tip). Plasticity implies on the one hand that the tip is
blunted, and on the other hand that inside the plastic zone the
Huber-von Mises criterion (\ref{Mises}) is satisfied. The outer
boundary of the plastic zone will be called below the ``yield
curve", and in polar coordinates around the crack tip will be
denoted $R(\theta)$.

Whatever is the actual shape of the
blunted tip its boundary cannot support normal components of the stress. Together with Eq. (\ref{Mises})
this implies that on the crack interface
\begin{equation}
\sigma_{1} =  \sqrt{3}~\sigma_{_{\rm Y}}\!,~~~~~~\sigma_{2} = 0.
\label{PS}
\end{equation}
On the other hand, the linear-elastic solution, which is still
valid outside the plastic zone, imposes the outer boundary
conditions on the yield curve. Below we will compute the outer
stress field {\em exactly} for an arbitrarily shaped crack using
the recently developed method of iterated conformal mappings
\cite{03BMP}. For the present argument we will take the outer
stress field to conform with the universal linear-elastic stress
field for mode I symmetry,
\begin{equation}
\sigma_{ij}(r,\theta) =  \frac{K_I}{\sqrt{2\pi r}} \Sigma^I_{ij}(\theta),
\label{UF}
\end{equation}
where for a crack of length $L$ with $\sigma^{\infty}$ being the
tensile load at infinity, the stress intensity factor $K_I$ is
expected to scale like $K_I\sim\sigma^\infty \sqrt{L}$. Using this
field we can find the yield curve $R(\theta)$. Typical yield curves
for straight and curved cracks are shown in the insets of Figs. \ref{straightsteps}
and \ref{upwardsteps}.

\begin{figure}[top]
\centering \epsfig{width=.45\textwidth,file=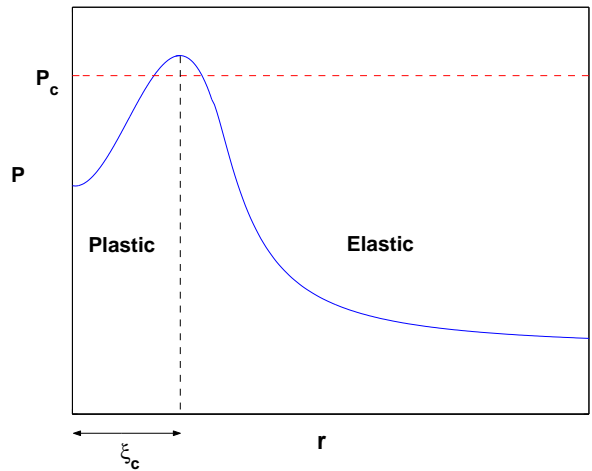}
\caption{A forward direction profile of the hydrostatic tension
$P$ in units of $\sigma_{_{\rm Y}}$. On the crack $P=\sqrt{3}/2$
and it attains a maximum of $\sqrt{3}$ on the yield curve. The
threshold line indicates a value of $P_c$ such that
$\case{\sqrt{3}}{2}\!\! <\! P_c \!\!<\!\! \sqrt{3}$. The typical
length $\xi_c$ is shown. Other directions exhibit qualitatively
similar profiles.} \label{StressProfile}
\end{figure}

\begin{figure}[here]
\centering \epsfig{width=.4\textwidth,file=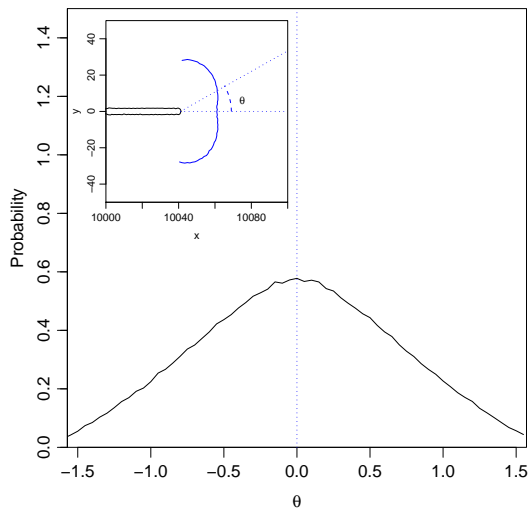}
\caption{The yield curve (inset) and probability distribution
function on it for a long straight crack. The distribution is
symmetric and wide enough to allow for deviations from the forward
direction.} \label{straightsteps}
\end{figure}

The typical scale $\xi_c$ follows from the physics of the
nucleation process. We assume that void nucleation occurs where
the hydrostatic tension $P$,  $P\equiv\case{1}{2}$Tr$\B \sigma$,
exceeds some threshold value $P_c$. Other assumptions on the
nature of the nucleation process will not affect qualitatively our
main result. The hydrostatic tension increases when we go away
from the tip and reaches a maximum near the yield curve. To see
this note that on the crack surface
$P=\case{\sqrt{3}}{2}\sigma_{_{\rm Y}}$ (cf. Eq. (\ref{PS})). On
the yield curve we use Eq. (\ref{UF}) and the Huber-von Mises
criterion together to solve the angular dependence of the
hydrostatic tension in units of $\sigma_{_{\rm Y}}$.  It attains a
maximal value of $\sqrt{3}\sigma_{_{\rm Y}}$ and is considerably
higher than $\case{\sqrt{3}}{2}\sigma_{_{\rm Y}}$ for a wide range
of angles. On the other hand the linear-elastic solution
(\ref{UF}) implies a monotonically decreasing $P$ outside the
yield curve. We thus expect $P$ to {\em attain its maximum value
near the yield curve}. This conclusion is fully supported by
Finite Element Method calculations, cf. \cite{85AM}. Finally,
since the nucleation occurs when $P$ exceeds a threshold $P_c$,
this threshold is between the limit values found above, i.e.
$\frac{\sqrt{3}}{2}\sigma_{_{\rm
Y}}\!\!\!<\!\!P_c\!\!<\!\!\sqrt{3}\sigma_{_{\rm Y}}$. The first
void will thus appear at a typical distance $\xi_c$ as shown in
Fig. \ref{StressProfile}. An immediate consequence of the above
discussion is that $\xi_c$ is related to the crack length via
\begin{equation}
\xi_c \sim \frac{K^2_I}{\sigma^2_{_{\rm Y}}} \sim \left(\frac{\sigma^{\infty}}{\sigma_{_{\rm Y}}}\right)^2
L \ .
\label{scaling}
\end{equation}
It is worthwhile to put this prediction to experimental test.
%%%%%%% FIGURE 2 %%%%%%%%%%%%%%%%%%
%%%%%%%%%%%%%%%%%%%%%%%%%%%%%%%%%

Naturally, the precise location of the nucleating void will
experience a high degree of stochasticity due to material
inhomogeneities. In our model below we will assume that the
nucleation can occur randomly anywhere in the region in which
$P\!>\!\!P_c$ with a probability proportional to $P\!-\!P_c$.

The simplest possible crack propagation model is obtained if we
assume that a void is nucleated and then the crack coalesces with
the void before a new void is nucleated. In experiments it appears
that several voids may nucleate before the coalescence occurs \cite{99BP,03Cil},
but we will demonstrate that already a one void model induces
positive correlations between consecutive void nucleations,
leading eventually to an anomalous roughness exponent larger than
0.5. Clearly, even this simple model requires strong tools to
compute the stress field around an arbitrarily shaped crack, to
determine at each stage of growth the location of the yield curve
and nucleating randomly the next void according to the probability
distribution discussed above. In a recent work we have developed
precisely the necessary tool in the form of the method of iterated
conformal mappings \cite{03BMP}.

In the method of iterated conformal mappings one starts with a
crack for which the conformal map from the exterior of the unit
circle to the exterior of the crack is known. (Below we start with
a long crack, in the form of a mathematical branch-cut of length
1000 in units of $\xi_c$). We can then grow the crack by little
steps in desired directions, computing at all times the conformal
map from the exterior of the unit circle to the exterior of the
resulting crack. Having the conformal map makes the {\bf exact}
calculation of the stress field (for arbitrary loads at infinity)
straightforward in principle and highly affordable in practice.
The details of the method and its machine implementations are
described in full detail in \cite{03BMP}. We should just stress
here that the method naturally grows cracks with finite curvature
tips, and each step adds on a small addition to the tip, also of a
finite size that is controlled in the algorithm.

Having the stress field around the crack we can readily find the
yield curve, and the physical region in its vicinity where a void
can be nucleated (naturally, the width of this region depends on
the critical value $P_c$ which is a parameter of the algorithm, as
$\sigma_{_{\rm Y}}$ is). Choosing with probability $\propto
P\!-\!P_c$ the position of the next void, we use this site as a
pointer that directs the crack tip. Fig. \ref{straightsteps} shows
a typical yield curve and the corresponding probability
distribution function ($ \propto P-P_c$) on this curve for a
straight crack. The distribution is symmetric and wide enough to
allow for deviations from the forward direction.

%%%%%%% FIGURE 3 %%%%%%%%%%%%%%%%%%
%%%%%%%%%%%%%%%%%%%%%%%%%%%%%%%%%%
We then use the method of iterated conformal mappings to make a
growth step to coalesce the tip with the void. Naturally the step
sizes are of the order of $\xi_c$. Thus the radius of curvature at
the tip is also of the order of $\xi_c$. We note that this model
forsakes the details of the void structure and all the
lengthscales below $\xi_c$. This is clearly acceptable as long as
we are mainly interested in the scaling properties on scales
larger than $\xi_c$.

%%%%%%% FIGURE 4 %%%%%%%%%%%%%%%%%%
\begin{figure}[top]
\centering \epsfig{width=.35\textwidth,file=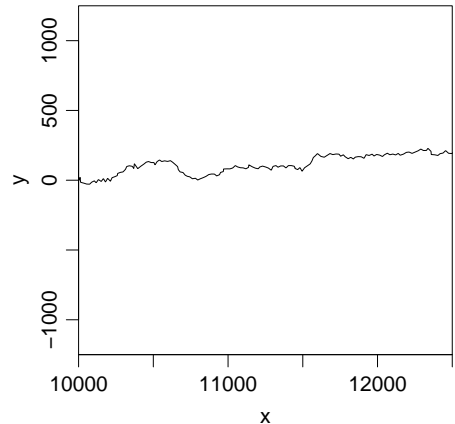}
\caption{A crack that was generated using our model.crack}
\label{crack}
\end{figure}
%%%%%%%%%%%%%%%%%%%%%%%%%%%%%%%%%%%

\begin{figure}[here]
\centering
\epsfig{width=.4\textwidth,file=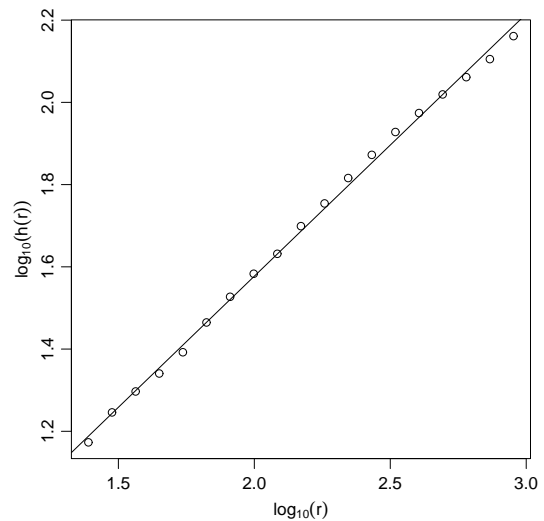}
\caption{Calculation of the anomalous roughening exponent.}
\label{h(r)}
\end{figure}

\begin{figure}[here]
\centering 
\epsfig{width=.4\textwidth,file=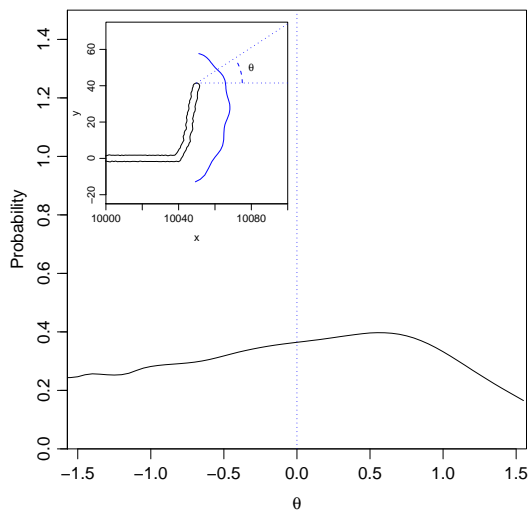} 
\caption{The yield curve (inset) and probability distribution function on it
for a straight crack followed by an upturn. The angle $\theta$ is measured relative to the
horizontal direction. It is clear that the distribution is skewed
in favor of positive angles.} \label{upwardsteps}
\end{figure}

In Fig. \ref{crack} we present a typical crack that had been grown
using this method. The positive correlation  between successive
void nucleation and coalescence events is obvious even to the
naked eye. Once the crack steps upward, there is a high
probability to continue upward, and vice versa. This is precisely
the property that we were after. A quantitative measurement of
this tendency is the roughening exponent, that we compute as
follows. Measuring the height fluctuations $y(x)$ in the graph of
the crack, one defines $h(r)$ according to
\begin{equation}
h(r)\equiv\left<Max\left\{y(\tilde x)\right\}_{x<\tilde x<x+r}-
Min\left\{y(\tilde x)\right\}_{x<\tilde x<x+r}\right>_x \ .
\label{rough}
\end{equation}
For self-affine graphs the scaling exponent $\zeta$ is defined via the scaling relation
\begin{equation}
h(r) \sim r^{\zeta} \ .
\end{equation}
In Fig. \ref{h(r)} we present a log-log plot of $h(r)$ vs. $r$,
with a best power-law fit of $\zeta=0.64\pm 0.04$. Indeed as
anticipated from the visual observation of Fig. \ref{crack} the
exponent is higher than 0.5.

%%%%%%% FIGURE 5 %%%%%%%%%%%%%%%%%%
%%%%%%%%%%%%%%%%%%%%%%%%%%%%%%%%%%

We note that our measured scaling exponent is still smaller than
the experimental one which is around 0.8. First let us say right
away that we do not expect in a 2-dimensional theory to hit a
3-dimensional exponent; scaling exponents are usually strongly
$d$-dependent even when the physics is invariant to the
dimensionality of space. In addition, we expect that a more
detailed model which incorporates a simultaneous multi-void
nucleation and coalescence would {\em increase} the positive
correlation in the positions of consecutive voids, and thus would
reduce the roughness of the surface (increase the scaling
exponent). 

The main points of the model are nevertheless worth
reiterating. First, we have a new typical scale, $\xi_c$, which is
crucial. Growing directly at the tip of the crack results in a
very strong preference for the forward direction, meaning that a step
up will most likely be followed by a step down, and vice versa, as
shown in \cite{02BLP}.
%%%%%%% FIGURE 6 %%%%%%%%%%%%%%%%%%
%%%%%%%%%%%%%%%%%%%%%%%%%%%%%%%%%%
The introduction of the physics
of the plastic zone results in creating a finite distance away
from the tip to realize the next growth step. Second, a growth in the
upward (or downward) direction is
affecting the next stress field such as to bias the next growth
step to be correlated with the last one. To see this clearly we
present in Fig. \ref{upwardsteps} the yield curve and the
corresponding probability distribution function ($ \propto P-P_c$)
on this curve for a long straight crack followed by an upward turn. It is clear that the distribution
is skewed in favor of positive angles with respect to the forward
direction. When the crack grows further this tendency becomes more
pronounced. This is the essence of the positive correlation
mechanism.

To improve our model further one needs to solve exactly for the stress field
around a crack and a single void ahead. This will allow the introduction of
two voids in the physically required places.  Such an improved model, which
is presently under construction, calls for mapping conformally doubly connected regions;
we expect such a model to lead to stronger positive correlations between growth steps
and to a higher scaling exponent.


\begin{thebibliography}{99}

\bibitem{84MPP}
B.B. Mandelbrot, D.E. Passoja and A.J. Paullay, Nature {\bf 308}, 721 (1984).

\bibitem{97B}
E. Bouchaud, J. Phys. Condens. Matter {\bf 9}, 4319 (1997).

\bibitem{03C}
F. C\'elari\'e, S. Prades, D. Bonamy, L. Ferrero, E. Bouchaud, C. Guillot and C. Marli\`ere, Phys.
Rev. Lett.
\bf{90}\rm, 075504 (2003).

\bibitem{Feder}
J. Feder, {\em Fractals}, (Plenum, New York, 1988).

\bibitem{97REF}
S. Ramanathan, D. Ertas, and D. S. Fisher
    Phys. Rev. Lett. {\bf 79}, 873 (1997).

\bibitem{02BLP}
F. Barra, A. Levermann and I. Procaccia, Phys. Rev. E {\bf66},   066122 (2002).


\bibitem{99BP}
E. Bouchaud and F. Paun, Comput. Sci. Eng., {\bf September/October}, 32 (1999).

\bibitem{03Cil}
S. Ciliberto, Private Communication, December 2003.

\bibitem{90Lub}
J. Lubliner, {\em Plasticity Theory}, (Macmillan, New York, 1990).

\bibitem{03BMP}
E. Bouchbinder, J. Mathiessen and I. Procaccia, ``Stress field
around arbitrarily shaped cracks in two-dimensional elastic
materials", Phys. Rev. E, In press. Also:cond-mat/0309523.
\bibitem{85AM}
N. Aravas and R.M. McMeeking, J. Mech. Phys. Solids {\bf33},  25 (1985).



\end{thebibliography}
\end{document}